\documentclass[aps,preprintnumbers, superscriptaddress, amsmath]{revtex4}

\begin{document}

\title{MAGNETISM IN COMPACT ULTRA DENSE OBJECTS}

\author{Efrain J. Ferrer\\\textit{Department of Physics, University of Texas at El Paso, El Paso, Texas
79968, USA\\ E-mail: ejferrer@utep.edu\\ www.utep.edu}}

\begin{abstract} This paper summarizes some of the recent results on magnetism in high dense mediums, where the phenomenon of color superconductivity
can be present, and its possible implications for the astrophysics of compact objects. The presentation will be organized through the answers to three
fundamental questions. \end{abstract}


\maketitle

\section{Can a Magnetic Field Modify the Color Superconducting  Pairing Pattern?}\label{aba:sec1} In spin-0 color superconductivity (CS), the
electromagnetism is not the conventional one. In the color superconducting medium the conventional electromagnetic field is not a real eigenfield.
Instead, in the presence of the diquark ground state the conventional electromagnetic field mixes with one gluon field leading to two new mass
eigenmodes. The massless resultant eigenmode is the new electromagnetic field in the medium, which is often called the rotated electromagnetic field.
Hence, even though the original electromagnetic $U(1)_{em}$ symmetry is broken by the formation of the electrically charged quark Cooper pairs, a
residual $\widetilde{U}(1)$ symmetry still remains. The massless gauge field associated with this symmetry in the color-flavor-locked (CFL) phase
\cite{CFL} of CS, for example, is given by the linear combination of the conventional photon field and the $8^{th}$ gluon field \cite{CFL, ABR},
${\tilde A}_\mu=\cos\theta A_\mu-\sin\theta G^8_\mu.$ Since the Cooper pairs are neutral with respect to the rotated electric charge, there is no
Meissner effect for the rotated electromagnetism.

The fact that a rotated magnetic field can penetrate the spin-0 color superconductor brings the possibility of looking for possible magnetic
field-interaction effects on the CS phase. One important consequence of the interaction with a strong magnetic field was first studied in
Ref. \cite{MCFL} in the context of the CFL phase. Although the Cooper pairs have zero net rotated charge, they could be formed by quarks of
opposite charges, as well as by neutral quarks. If the magnetic field is strong enough so that the magnetic length $l_0=1/\sqrt{2eB}$ becomes smaller
than the pairs' coherence length, then the magnetic field can interact with the pair constituents and significantly modify the pair structure of the
condensate. As shown in Refs. \cite{MCFL}, the presence of a magnetic field changes the CFL phase from having a simple gap, to a new phase with
three different gaps. The new phase that forms in the presence of the magnetic field also has color-flavor-locking, but with a smaller symmetry group
$SU(2)_{C+L+R}$, a change that is reflected in the splitting of the gaps. The realization of the CFL pairing that occurs in the presence of a magnetic
field, is known as the magnetic-CFL (MCFL) phase \cite{MCFL}. The MCFL phase has similarities, but also important differences with the CFL phase
\cite{MCFL, MCFLoscillation, phases}.

\section{Can a Magnetic Field be Modified by a Color Superconductor?} Another nontrivial electromagnetic effect in CS is that some of the gluons
become charged, hence they can interact with the rotated magnetic field.

Taking into account the Schwinger rest-energy spectrum of a charged particle of spin $s$, charge $q$, gyromagnetic ratio $g$, and mass $m$ in a
magnetic field $H$, \begin{equation} \label{Rest-Energy} E^2_n=(2+1)qH-gq \textbf{H}\cdot\textbf{s}+m^2, \end{equation} we see that for spin-1
particles (i.e. g = 2 and spin projections -1, 0, +1) one of the modes of the charged gauge field becomes tachyonic ($E^2 < 0$) for strong enough
magnetic fields ($H > Hcr = m^2/e$). This is the well known "zero-mode problem" found in different QFT contexts in the presence of a magnetic field
\cite{zero-mode}. Then, once the gluons can be affected by an applied magnetic field in a CS state, an instability arises at sufficiently high field
strengths through this mechanism. To remove the field-induced instability, a vortex state, characterized by the condensation of charged gluons and the
creation of magnetic flux tubes, is formed \cite{Vortex}. Inside the flux tubes the magnetic field is stronger than the applied one.

The origin of the anti-screening effect \cite{Vortex} follows from the minimum equations for the interacting gluon field $\overline{G}$ and the
induced rotated magnetic field $\widetilde{B}$ in the presence of the applied field $\widetilde{H}$ \begin{equation} \label{EqG} \widetilde{\Pi}^{2}
\overline{G}+2(m_{M}^{2}-\widetilde{e}\widetilde{B})\overline{G}+8g^{2}\overline{G}^{2}\overline{G}=0, \end{equation}

\begin{equation} \label{EqB} 2\widetilde{e} \overline{G}^{2}-\widetilde{B}+\widetilde{H}=0 \end{equation} Identifying $\overline{G}$ with the complex
order parameter, Eqs. (\ref{EqG})-(\ref{EqB}) become analogous to the Ginzburg-Landau equations for a conventional superconductor except the negative
sign in front of the $\widetilde{B}$ field in Eq. (\ref{EqG}) and the positive sign in the first term of the LHS of Eq. (\ref{EqB}) \cite{Vortex}. The
fact that those signs turn to be opposite to those appearing in conventional superconductivity is due to the different nature of the condensates in
each case. While in conventional superconductivity the Cooper pair is formed by spin-1/2 particles, here we have a condensate formed by spin-1 charged
particles interacting through their anomalous magnetic moment with the magnetic field.

Notice that because of the different sign in the first term of (\ref{EqB}), contrary to what occurs in conventional superconductivity, the resultant
field $\widetilde{B}$ is stronger than the applied field $\widetilde{H}$. Thus, when a gluon condensate develops, the magnetic field will be
antiscreened and the color superconductor will behave as a paramagnet. We conclude that the magnetic field in the new phase is boosted to a higher
value, which depends on the modulus of the $\overline{G}$-condensate. Thus, the vortex solution gives rise to a new phase, that is known as the
Paramagnetic-CFL (PCFL) phase \cite{Vortex}.

\section{Does Magnetic Color Superconductivity have any Astrophysical Relevance?} An important characteristic of neutron stars is that they typically
possess very strong magnetic fields, which are estimated to reach values up to $10^{20}$ G in their inner core if formed by quark matter \cite{EoS}.
Unveiling the interconnection between the star's magnetic field and its dense phases is important to understand the interplay between QCD and neutron
star phenomenology. Among the possible consequences of the existence of the MCFL phase in the inner core, one can mention, for example, the effect on
the star's transport properties, which are determined by the low-energy spectrum of the phase \cite{phases}. Also, the possibility to trigger the
magnetic generation of vortices of charged gluons in the PCFL phase can serve as a mechanism to enhance the existing magnetic field, without having to
rely only on the quick spinning assumed in the standard model of magnetars that, as known, are associated with some of the conflict of this model with
observations. On the other hand, strong magnetic fields can affect the stability of quark stars. That is, depending on whether the energy density of
the MCFL phase is higher or smaller than that of the iron nucleus ($\sim 930 MeV$), the content of a magnetized strange quark could be or not made of
MCFL matter. For more details on these topics see Ref. \cite{Review}.

\begin{acknowledgments} 
This work has been supported in part by DOE Nuclear Theory grant DE-SC0002179. 
\end{acknowledgments}

\end{document}